\documentclass[%
groupedaddress,
unsortedaddress,
amsmath,amssymb,
aps,
]{revtex4-2}

\pdfoutput=1

\usepackage[english]{babel}
\usepackage[utf8]{inputenc}
\usepackage{graphicx}% Include figure files
\usepackage{dcolumn}% Align table columns on decimal point
\usepackage{bm}% bold math
\usepackage{hyperref}
\usepackage{upgreek}
\usepackage[table]{xcolor}
\usepackage{setspace}

\usepackage{array}
\newcolumntype{P}[1]{>{\raggedright\arraybackslash}p{#1}}

\setcitestyle{comma,super}

\begin{document}

%\preprint{APS/123-QED}

\title{Plasma-MDS, a metadata schema for plasma science with examples from plasma technology}

\author{Steffen Franke}
 \affiliation{%
Leibniz Institute for Plasma Science and Technology (INP), Felix-Hausdorff-Str. 2, 17489 Greifswald, Germany
}%Lines break automatically or can be forced with \\

\author{Lucian Paulet}%
\affiliation{%
Leibniz Institute for Plasma Science and Technology (INP), Felix-Hausdorff-Str. 2, 17489 Greifswald, Germany
}%

\author{Jan Schäfer}%
\affiliation{%
Leibniz Institute for Plasma Science and Technology (INP), Felix-Hausdorff-Str. 2, 17489 Greifswald, Germany
}%

\author{Deborah O'Connell}%
\affiliation{%
York Plasma Institute, Department of Physics, University of York, Heslington, York, YO10 5DD, U.K
}%

\author{Markus M. Becker}%
\email{markus.becker@inp-greifswald.de}
\affiliation{%
Leibniz Institute for Plasma Science and Technology (INP), Felix-Hausdorff-Str. 2, 17489 Greifswald, Germany
}%

\date{\today}% It is always \today, today,

\begin{abstract}
	A metadata schema, named Plasma-MDS, is introduced to support research data management in plasma science.
	Plasma-MDS is suitable to facilitate the publication of research data following the FAIR principles in domain-specific repositories and with this the reuse of research data for data driven plasma science. 
	In accordance with common features in plasma science and technology, the metadata schema bases on the concept to separately describe the source generating the plasma, the medium in which the plasma is operated in, the target the plasma is acting on, and the diagnostics used for investigation of the process under consideration. These four basic schema elements are supplemented by a schema element with various attributes for description of the resources, i.e. the digital data obtained by the applied diagnostic procedures. 
	This novel metadata schema is first applied for the annotation of datasets published in INPTDAT---the interdisciplinary data platform for plasma technology.
\end{abstract}

\maketitle

%\tableofcontents

%\section*{Comments}

%\texttt{The ‘Article’ format can be used to present original reports on systems or techniques that clearly advance data sharing and reuse to support reproducible research. This includes research on sharing, managing and processing scientific research data. Articles describing data repositories, standards and ontologies are welcome when they include compelling demonstrations of data exchange, enrichment or knowledge generation made possible by the system or standard. Articles may also present sociological research on data sharing habits or perceptions, or the effectiveness of sharing policies. Computer science or engineering articles describing data processing or analysis techniques may be considered in some cases, when they would be relevant to a broad segment of the journal’s audience.}

%\texttt{For Analyses and Articles, the main text (excluding abstract, Methods, references and figure legends) is approximately 3,000 words. The abstract is typically 100-170 words, unreferenced. An Introduction is followed by sections headed Results, Discussion, Methods. All Analysis and Article submissions should include a data availability statement. Guidance on writing a data availability statement can be found here. The Methods and Results should be divided by topical subheadings; the Discussion does not contain subheadings. The Methods should be followed by References, Acknowledgements and a Competing interests statement.}

\section*{Introduction}

The rapid progress in data science methods for machine-based analysis of big data provides enormous potential for new data driven sciences and the development and optimization of innovative technologies. 
In the wide field of plasma science, the application of machine learning methods, e.g. for investigation and control of fusion plasmas, the particle and event identification in high energy physics, and the discovery of space phenomena in astrophysics has been common practice for several years, see \cite{Kates-Harbeck2019,Albertsson_2018,Spears-2018} and references therein.
Recently, first approaches have been published that use machine learning methods for simulation, diagnostics and control of technological plasmas~\cite{Mesbah2019,Kruger-2019-ID5326}. This is of particular interest because technological plasmas are used in many applications and industrial processes. Examples are the deposition of thin films, plasma etching, and plasma decontamination~\cite{Brandenburg-2018-ID5134,Cvelbar-2018-ID5129,Simek2019}. During the last ten years, plasma medicine has been established as an additional important research topic in the field of cold plasmas and first certified medical devices are already in practical use~\cite{vonWoedtke-2013-ID3925,Weltmann2016}. Applications of cold plasmas in medicine include the plasma-based synthesis of biomedical surfaces, wound healing, and cancer treatment~\cite{Bekeschus-2018-ID5112}.

The potential of data driven science in plasma physics---like in all other fields---can only be fully explored if research data is findable, accessible, interoperable and reusable (FAIR) for both humans and computers.
This requirement has recently been pinpointed  by the FAIR data principles~\cite{Wilkinson2016} (cf. Table~\ref{table:FAIR}). The minimum requirements for ``fair'' research data are that the data is made public and that it is well documented by additional metadata.
The quality of metadata plays a key role for the degree of ``fairness''. Once the (meta)data is registered or indexed in a searchable resource with a unique and persistent identifier, the machine-readable metadata should contain information on how the data can be accessed, how it can interoperate with applications or workflows for analysis, storage and processing, and in which context it can be reused, i.e., detailed information on the scope of the data, lab conditions, process parameters, \textit{etc.}~\cite{Chen2019}.

%\begin{wraptable}{r}{9.5cm}
\begin{table}[h]
	\caption{The FAIR Guiding Principles according to Wilkinson \textit{et al.} \cite{Wilkinson2016}.}
	\label{table:FAIR}    
	\small
	\centering
	\begin{tabular}{|l|p{12cm}|}
		\hline
		\multicolumn{2}{|l|}{\it To be \textbf{F}indable} \\
		\hline
		F1 & (meta)data are assigned a globally unique and persistent identifier \\\hline
		F2 & data are described with rich metadata (defined by R1 below) \\\hline
		F3 & metadata clearly and explicitly include the identifier of the data it describes \\\hline
		F4 & (meta)data are registered or indexed in a searchable resource \\\hline
		\hline
		\multicolumn{2}{|l|}{\it To be \textbf{A}ccessible} \\
		\hline
		A1 & (meta)data are retrievable by their identifier using a standardized communications protocol \\\hline
		A1.1 & the protocol is open, free, and universally implementable \\\hline
		A1.2 & the protocol allows for an authentication and authorization procedure, where necessary \\\hline
		A2 & metadata are accessible, even when the data are no longer available\\\hline
		\hline
		\multicolumn{2}{|l|}{\it To be \textbf{I}nteroperable} \\
		\hline
		I1 &  (meta)data use a formal, accessible, shared, and broadly applicable language for knowledge representation \\\hline
		I2 & (meta)data use vocabularies that follow FAIR principles \\\hline
		I3 & (meta)data include qualified references to other (meta)data \\\hline
		\hline
		\multicolumn{2}{|l|}{\it To be \textbf{R}eusable} \\
		\hline
		R1 & meta(data) are richly described with a plurality of accurate and relevant attributes \\\hline        
		R1.1 & (meta)data are released with a clear and accessible data usage license \\\hline
		R1.2 & (meta)data are associated with detailed provenance \\\hline
		R1.3 & (meta)data meet domain-relevant community standards \\
		\hline                
	\end{tabular}
\end{table}
%\end{wraptable}

In many scientific disciplines established scope-specific metadata standards exist that are recognised and broadly used by the community. To give an example, the data tag suite DATS has recently been introduced to enable discoverability of datasets in the field of biomedical research~\cite{Sansone2017}.
Dictionaries of disciplinary metadata standards are provided by the Digital Curation Centre (\url{http://www.dcc.ac.uk/resources/metadata-standards/}, accessed: 2020-06-14) and the Research Data Alliance (\url{http://rd-alliance.github.io/metadata-directory/}, accessed: 2020-06-14), for example. 
In some cases, elaborated data portals are already available that provide public access to data for reuse in data driven research. 
Examples are the proteomics identifications (PRIDE) database~\cite{pride}, 
the novel materials discovery (NOMAD) repository~\cite{Ghiringhelli2017},
%the CHRS Data Portal for global satellite precipitation data~\cite{Nguyen2019}, 
and the repository for high energy physics data (HEPData)~\cite{Maguire_2017}.
Although there are a number of databases which are relevant for research in applied plasma science and technology, e.g. the NIST Atomic Spectra Database (\url{https://www.nist.gov/pml/atomic-spectra-database}, accessed: 2020-06-14), the LXcat database~\cite{Pitchford-2017-ID4155}, %\cite{lxcat},
and Quantemol-DB~\cite{Tennyson-2017-ID4197}, %\cite{qdb}
%or the InSpire database driven by CERN, DESY, Fermilab and SLAC (http://inspirehep.net/info/general/project/index?ln=de) [better example?], 
there is no repository that is specifically designed for the curation of the heterogeneous data from research in the field of applied plasma physics. This hinders the reuse of and access to data in this specific domain. 
%(cf. table~\ref{table:FAIR}). 
%Furthermore, there are no metadata standards to describe this kind of research data efficiently. However, domain-specific metadata are required to enable findability, interoperability, and reusability of data (cf. table~\ref{table:FAIR}). 
And even more important, there are no metadata standards for a unified categorisation and detailed description of research data in plasma physics. 
It is important to note that certain data models exist, which mainly aim at a uniform storage of rather homogeneous data and with this strongly support the interoperability of data in specific areas.
Examples of such physical data models in plasma related sciences include the data model of the HEPData platform~\cite{Maguire_2017}, the ITER Physics Data Model~\cite{Imbeaux_2015}, and the data acquisition and analysis system, MDSplus~\cite{Fredian-2018}.
However those data models are more or less specfic to the resources deposited in according databases and do not allow for a more general categorisation and findability of research data as it is intended with the proposed metadata schema.

The present manuscript suggests a metadata schema for research data in plasma science and technology, which is named Plasma-MDS.
Plasma-MDS is supposed to be complementary to established data models and aims to be a starting point for the development of a standard for the categorisation and documentation of digital data obtained from research in plasma physics.
With this, it supports recent attempts to enable ``a new era of plasma science and technology research and development'' (\url{https://www.york.ac.uk/physics/ypi/conferencesevents/icddps/}, accessed: 2020-06-14) by data-driven discovery in plasma science and provides a basis for participation of the community in comprehensive developments with respect to research data management, like for example the European Open Science Cloud (EOSC) (\url{https://www.eosc-portal.eu}, accessed: 2020-06-14) and the National Research Data Infrastructure (NFDI) in Germany~\cite{nfdi}. 
Examples of using Plasma-MDS for the categorisation and description of datasets from research in plasma technology are given in the present manuscript.

%The manuscript is organized as follows. In section 2 the metadata schema Plasma-MDS  is introduced. Then, its practical application is demonstrated using two examples in section 3 and in section 4 the benefits of Plasma-MDS with respect to the FAIR data principles are discussed. Section 5 summarizes the work and outlines necessary future steps.

%\clearpage
\section*{Results}

Metadata represent extra information attached to data that allows people and automated processes to find, access and ultimately reuse data. Among others, Dublin Core (\url{http://dublincore.org/schemas/}, accessed: 2020-06-14), the \mbox{DataCite} Metadata Schema~\cite{datacitems}, and DCAT (\url{https://www.w3.org/TR/vocab-dcat-2/}, accessed: 2020-06-14) represent fundamental metadata schemata that are widely used for the collection and indexing of general metadata of digital objects, such as title, publication year, and permanent identifier. However, the ``degree of fairness'' of public research data (cf. table~\ref{table:FAIR}) can dramatically be enhanced by adding additional domain-specific metadata. 

The new metadata schema Plasma-MDS can be used as an extension to general metadata schemata in plasma science and technology.
It follows the nomenclature ``schema.element. qualifier'' and comprises various metadata fields related to the \textit{plasma source}, the \textit{plasma medium}, and the \textit{plasma target} possibly involved in the study. Furthermore, metadata fields related to the applied \textit{diagnostics} and the published \textit{resource} are included. Here, the schema element diagnostics also aims to cover applied modelling and simulation methods.
The motivation for the main schema elements ``source'', ``medium'', ``target'', and ``diagnostics'' lies in the fact that scientific results in plasma physics frequently refer to a plasma source (e.g. atmospheric pressure plasma jet) which is operated in a medium (e.g. argon) and acting on a target (e.g. biological tissue). Furthermore, plasma physics utilizes a variety of diagnostic methods (e.g. laser absorption spectroscopy) and there are numerous scientific papers which concentrate on specific aspects of the plasma diagnostics rather than on a certain plasma (source, medium, and target). 
Therefore, plasma diagnostics is also considered as a separate schema element.
In addition, the schema element ``plasma.resource'' specifies  details of the digital data object, which is obtained by the applied diagnostic procedures.

Only the diagnostics and resource  metadata fields are designed to be mandatory. This is because in one study the focus might be on the diagnostics applied to a certain target, while in another the simulation of a plasma source without inclusion of any target might be of interest. However, it is strongly suggested to complete as many metadata fields as possible in order to ensure a high level of ``fairness'' of the data.
Furthermore, there is no controlled vocabulary so far to provide maximum flexibility in the definition of metadata. It is intended to review the plasma metadata schema after an initial phase of growing usage and to evaluate the establishment of a community standard including controlled vocabularies.

\begin{figure}[htb]\centering
	\includegraphics[width=.5\linewidth]{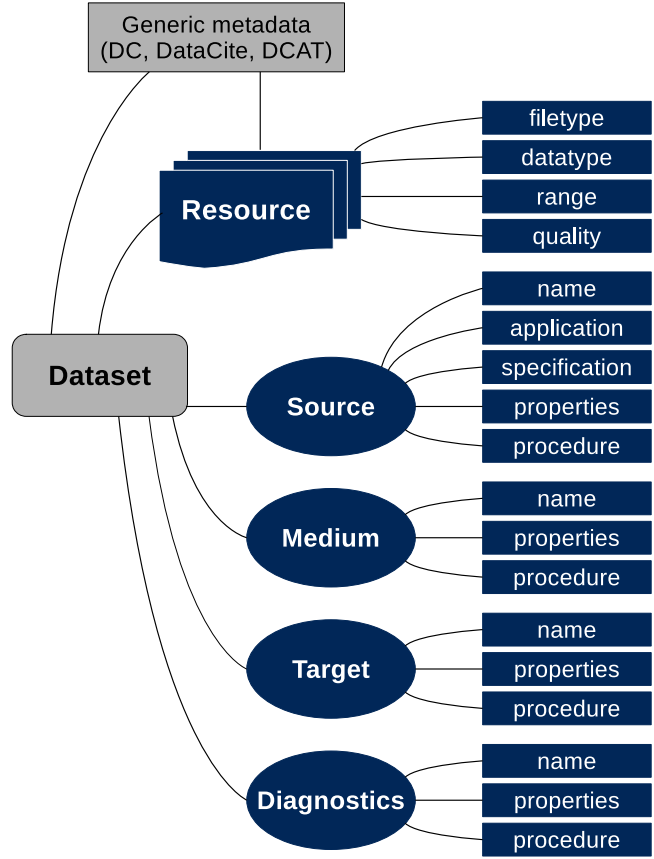}
	\caption{Overview of the schema elements and qualifiers of Plasma-MDS (blue). The sketch illustrates how the domain-specific schema extends general metadata of datasets according to basic metadata schemata like Dublin Core (DC), DataCite, or DCAT. \label{fig:schema}}
\end{figure}

An overview of the schema elements with the respective qualifiers is presented in figure~\ref{fig:schema}. Here, it is also shown how Plasma-MDS is used as a domain-specific extension to general metadata schemata such as Dublin Core, DataCite, or DCAT. This means that Plasma-MDS does not aim to replace existing standards, but it adds the option to describe studies conducted in the field of plasma physics in more detail.
Note that the specific sub-domain and/or topic (e.g. inertial confinement fusion, low-temperature plasmas, plasma medicine) to which the record refers should be named by the general schema used in the respective case, e.g. in ``\href{http://purl.org/dc/terms/subject}{dcterms.subject}''. Moreover, data models, which may be used to store the digital resources should also be referenced by the general schemata, e.g. ``\href{http://purl.org/dc/terms/conformsTo}{dcterms.conformsto}''. 
%It must be emphasized that a data model, like ITER Physics Data Model~\cite{Imbeaux_2015}, should be clearly distinguished from a metadata schema, like Plasma-MDS. Plasma-MDS is intended to categorise datasets containing resources which might conform to different data models.  

Plasma-MDS distinguishes the three different field types ``controlled list'', ``term list'', and ``free text''. 
Controlled list means that pre-defined categories are available for selection.
Term lists are defined as compilations of keywords generated on-the-fly from terms already describing the respective element in a specific data repository using Plasma-MDS. 
They are used whenever the establishment of controlled lists maintained by the community is an option in the long-term perspective. On the other hand, free text fields aim to give more detailed information on the respective element that cannot be represented by well-defined terms.

\subsection*{Schema element plasma.source}

The schema element ``plasma.source'' has five qualifiers. First, ``plasma.source.name'' designates the plasma source. Several plasma source names can be entered here, if the data are related to several plasma sources. Furthermore it might be helpful to give not only the trademark name (e.g. ``kINPen\textsuperscript{\textregistered} MED'', which is a certain plasma source developed at the INP) but to name also the type of the plasma source (e.g. ``HF plasma jet'', which indicates that the ``kINPen\textsuperscript{\textregistered} MED'' is a high frequency plasma jet). This should increase findability of datasets including sought-for plasma sources. Before adding a new plasma source name to a database using Plasma-MDS as metadata schema, it should be verified that this value is not already given taking care for differing notations. 

Next, the qualifier ``plasma.source.application'' informs about the application area the plasma source and the dataset are related to. Several terms can be given here describing different aspects of the application, for example plasma medicine or surface treatment. The first term identifies the dataset to be related to the topic plasma medicine, whereas the second term describes the more technical aspect of surface treatment in contrast, e.g. to plasma (volume) chemistry. Terms like ``antimicrobial reduction'' indicate the purpose of the application, if it should be distinguished from others, e.g. ``modification of wettability''.

The qualifier ``plasma.source.specification'' allows to define basic specifications of the plasma source, which are i) current/voltage waveform, ii) frequency range, iii) pressure range and iv) temperature range. These four specifications describe basic properties which can be applied to almost every plasma source and ensure a rough categorisation of the plasma:
\begin{enumerate}
	\item ``waveform'' specifies the power delivery waveform and can take the values ``pulsed'', ``DC'' (direct current), and ``AC'' (alternating current);
	
	\item ``frequency'' specifies the pulse repetition frequency or the frequency of the waveform, and can take the values ``low frequency'' ($<300$\,kHz), ``high frequency'' (300\,kHz to 300\,MHz), and ``microwave'' ($>300$\,MHz). No value is to be added if  ``waveform'' is set to ``DC'';
	
	\item ``pressure'' specifies the gas pressure and can take the values ``low pressure'' ($\lesssim 10^{3}$\,Pa), ``medium pressure'' ($10^3$ to $10^5$\,Pa), ``atmospheric pressure'' ($\approx10^5$\,Pa), and ``high pressure'' ($\gtrsim 10^5$\,Pa);
	
	\item ``temperature'' specifies the state of thermodynamic equilibrium and can take the values ``thermal'' and ``non-thermal'', which are fundamental categories to describe if a plasma is in local thermal equilibrium or not. 
\end{enumerate}

With the qualifier ``plasma.source.properties'' it is possible to add further description of plasma properties as free text. Finally, ``plasma.source.procedure'' is a free text container to describe procedures to set the plasma source into operation. But it can also be used to give details on the whole (experimental) setup needed to produce the data resource. 
Table~\ref{table:Plasma-MDS.src} gives an  overview over all qualifiers of the schema element ``plasma.source''.

%\begin{threeparttable}[ht]
\begin{table}[ht]
	\caption{Plasma-MDS fields related to the plasma source.}
	\label{table:Plasma-MDS.src}
	\small
	\centering   
	\begin{tabular}{ |l|P{2.5cm}|P{5cm}|P{1.5cm}|P{4cm}|}
		\hline
		\textit{\textbf{Id}} 
		& \textit{\textbf{Field}}         
		& \textit{\textbf{Definition}} 
		& \textit{\textbf{Format}}
		& \textit{\textbf{Example}} \\
		\hline
		1.1 & plasma.source. name 
		& name and/or type of the plasma source
		& term list
		&  kINPen\textsuperscript{\textregistered} MED, COST jet, HF plasma jet\\
		\hline
		1.2 & plasma.source. application
		& application the plasma source is applied for 
		& term list
		& plasma medicine, surface treatment, antimicrobial reduction\\
		\hline
		1.3 & plasma.source. specification
		& technical specifications of the plasma source (waveform, frequency, pressure, temperature)
		& controlled list
		& AC, high frequency, low pressure, non-thermal\\
		\hline
		1.4 & plasma.source. properties
		& properties of the plasma source
		& free text
		& details on power input, current/voltage amplitude \textit{etc.}\\              
		\hline
		1.5 & plasma.source. procedure
		& procedure to prepare the plasma source; this field should be used to described the whole procedure, including medium and target
		& free text
		& e.g. temperature conditioning for each parameter set\\           
		\hline
	\end{tabular}
	%\begin{tablenotes}
	%    \item[$\dagger$] Controlled lists will continuously be extended and reviewed according to the needs of the plasma physics community. 
	%\end{tablenotes}
\end{table}

The metadata collected by the schema element ``plasma.source'' are not mandatory and can be omitted if the dataset includes data which are not specific to a plasma source, e.g. data from target analysis. 
However, metadata for the plasma source should be included whenever applicable. For instance, the metadata of a plasma source used for pre-treatment of a specific target might also be included if the dataset contains only research data from target analysis.

\subsection*{Schema element plasma.medium}

The schema element ``plasma.medium'' has three qualifiers and describes the medium the plasma is operated in or consisting of. First, ``plasma.medium.name'' names the medium. Examples are noble gases (e.g. Ar), molecular gases (e.g. CO$_2$), or complex mixtures, e.g. plasma compositions consisting of sulfur hexfluoride (SF$_6$) and polytetrafluoroethylene (PTFE). Arc plasmas operated in vacuum usually consist of evaporated electrode material like copper and chromium (Cu-Cr).  For gas mixtures it is favourable to fill this field with a list of different species rather than to name each single mixture of species. Furthermore, chemical element symbols and common abbreviations are preferred, e.g. ``Ar'' instead of ``argon'' and ``HMDSO'' instead of ``hexamethyldisiloxane''.
Next, ``plasma.medium.properties'' is a free text qualifier that can take unstructured information to describe details of the plasma medium, e.g. gas flow rates, the carrier gas in a mixture, or the gas purity. Finally, ``plasma.medium.procedure'' is a free text container to describe procedures to prepare the medium before plasma operation and the treatment during plasma operation. Table~\ref{table:Plasma-MDS.medium} gives an overview over all qualifiers of the schema element ``plasma.medium''.

\begin{table}[ht]
	\caption{Plasma-MDS metadata fields related to the plasma medium.}
	\label{table:Plasma-MDS.medium}
	\small
	\centering   
	\begin{tabular}{ |l|P{2.5cm}|P{5cm}|P{1.5cm}|P{4cm}|}
		\hline
		\textit{\textbf{Id}} 
		& \textit{\textbf{Field}}         
		& \textit{\textbf{Definition}} 
		& \textit{\textbf{Format}}
		& \textit{\textbf{Example}} \\
		\hline
		2.1 & plasma.medium. name         
		& medium name the plasma source is operated in or acting on
		& term list 
		& Ar, CO$_2$, H$_2$O, air \\
		\hline
		2.2 & plasma.medium. properties       
		& properties of the medium the plasma source is operated in or acting on%, like carrier gas (mixture)
		& free text
		& gas flow rate: 100 sccm, carrier gas: Ar, precursor: HMDSO, gas mixture: Ar with 10 ppm HMDSO\\
		\hline
		2.3 & plasma.medium. procedure    
		& standard procedure to prepare the medium
		& free text
		& gas flow has to be established for at least 30\,s before plasma ignition.\\           
		\hline
	\end{tabular}
\end{table}

The schema element ``plasma.medium'' is not mandatory and can be omitted, e.g. if the description of the plasma source already provides sufficient information on the plasma medium. This might be the case, e.g. if the plasma source is a low-pressure sodium lamp, where the lamp fill is part of the plasma source specification. However, for reasons of findability redundant information on the plasma medium in the corresponding schema element is suggested.

\subsection*{Schema element plasma.target}

As for ``plasma.medium'', there are three qualifiers for the schema element ``plasma.target'' which allow to specify the name,  properties, and procedure of the target.
The qualifier ``plasma.target.name'' should designate the target the plasma source is acting on---either directly or mediated by a substance. Examples for possible target names are ``Si wafer'', ``distilled water'' and ``E. coli''.
It is suggested to use chemical element symbols or common abbreviations where applicable.
Multiple targets can be named here. This is of particular importance if the action of the plasma is mediated by a substance. For instance, it may be of interest to treat water or pharmaceuticals in a plasma reactor and afterwards use those treated substances to let them interact with a cell line. Such cases are considered by specifying multiple plasma targets. 

The qualifier ``plasma.target.properties'' is designed to collect details of the plasma target, e.g. geometric dimensions, grade, and orientation of a silicon wafer. Consequently, the qualifier ``plasma.target.procedure'' is eligible to describe any processing steps to prepare targets before plasma treatment (e.g. growth of cell lines) as well as handling throughout the plasma treatment. Table~\ref{table:Plasma-MDS.target} gives an overview over all qualifiers of the schema element ``plasma.target''.

\begin{table}[ht]
	\caption{Plasma-MDS metadata fields related to the plasma target.}
	\label{table:Plasma-MDS.target}
	\small
	\centering   
	\begin{tabular}{ |l|P{2.5cm}|P{5cm}|P{1.5cm}|P{4cm}|}
		\hline
		\textit{\textbf{Id}} 
		& \textit{\textbf{Field}}         
		& \textit{\textbf{Definition}} 
		& \textit{\textbf{Format}}
		& \textit{\textbf{Example}} \\
		\hline
		3.1 & plasma.target. name 
		& name of the target the plasma source is acting on, either directly or mediated by a medium
		& term list
		& Si waver, distilled water, Escherichia coli\\
		\hline
		3.2 & plasma.target. properties
		& properties of the target the plasma source is acting on
		& free text
		& silicon wafer: 100 mm diameter, prime grade, orientation 100, E. coli (DSM 11250, NCTC 10538)
		\\              
		\hline
		3.3 & plasma.target. procedure
		& standard procedure to prepare the target (pre-treatment) and handling throughout plasma treatment  
		& free text     
		& E. coli prepared on glass substrate according to internal procedure\\
		\hline
	\end{tabular}
\end{table}

The schema element ``plasma.target'' is not mandatory and can be omitted, e.g.  if only the characterization of a plasma source is intended.

\subsection*{Schema element plasma.diagnostics}

The schema element ``plasma.diagnostics'' serves the purpose to give details on the respective plasma diagnostics and modelling/simulation procedures. That means, it is about the methods used in the study to produce the data resource, either experimentally or theoretically. This is of particular importance as in plasma physics numerous specialized diagnostic methods are relevant and filtering datasets according to the applied diagnostics can be helpful. Another advantage of this schema element is that datasets can be considered which are related to plasma physics but do not deal with a specific plasma source or plasma application. For instance, this is the case if a diagnostic or modelling/simulation method is reported which is not only applicable to plasmas but also to non-ionized gases, i.e. vapours or cold gas. Examples of plasma diagnostics names include ``OES'' (optical emission spectroscopy), ``XPS'' (X-ray photoelectron spectroscopy), ``PIC-MCC'' (particle-in-cell/Monte Carlo collision simulations). It is suggested to use common abbreviations where available. However, synonyms can be entered here as well together with different diagnostics applied within one dataset.

The second and third qualifiers ``plasma.diagnostics.properties'' and ``plasma.diagnostics.procedure''  contain further details on the applied diagnostics and modelling/simulation methods, respectively. References to journal publications with more details on the applied methods may be provided here. Table~\ref{table:Plasma-MDS.diagnostics} gives an  overview of all qualifiers of the schema element ``plasma.diagnostics''.

\begin{table}[ht]
	\caption{Plasma-MDS metadata fields related to diagnostics.}
	\label{table:Plasma-MDS.diagnostics}
	\small
	\centering   
	\begin{tabular}{ |l|P{2.5cm}|P{5cm}|P{1.5cm}|P{4cm}|}
		\hline
		\textit{\textbf{Id}} 
		& \textit{\textbf{Field}}         
		& \textit{\textbf{Definition}} 
		& \textit{\textbf{Format}}
		& \textit{\textbf{Example}} \\
		\hline
		4.1 & plasma. \mbox{diagnostics}. \mbox{name}        
		& name of the applied diagnostics or modelling/simulation method
		& term list
		& OES, LAAS, XPS, SEM, PIC-MCC, fluid-Poisson model\\
		\hline
		4.2 & plasma. \mbox{diagnostics}. \mbox{properties}
		& properties of the applied diagnostics which are not part of the resource metadata      
		& free text  
		&  laser diode at 395\,nm and 50\,mW  \\
		\hline   
		4.3 & plasma. \mbox{diagnostics}. \mbox{procedure}
		& details of the applied diagnostic procedures which are not part of the resource metadata      
		& free text  
		&  voltage is measured between the contact tube of the welding torch and the workpiece\\
\hline  		     
	\end{tabular}
\end{table}

The schema element ``plasma.diagnostics'' is mandatory because knowledge of the applied experimental/modelling/simulation method is assumed to be  crucial for the reusability of the data.

\subsection*{Schema element plasma.resource}

The Plasma-MDS is designed to describe datasets which can contain several resources. Resources are digital representations of research data. Hence, the above defined metadata do not serve for the only purpose to describe a single resource but  possibly a set of resources. To give details on the specifics of each resource, the schema element ``plasma.resource'' is introduced. The qualifiers might be in parts redundant to metadata of different metadata schemata like, e.g. Dublin Core and DataCite Metadata Schema. However, they provide key information on each single resource which should be compiled into the schema element ``plasma.resource'' for the sake of clarity. The qualifiers are defined as follows:
\begin{enumerate}
	\item ``filetype'' obviously contains the file extension, e.g. pdf, csv or jpg. It corresponds to  ``\href{http://purl.org/dc/terms/format}{dcterms.format}'' and should be appropriate for long-term preservation;	

	\item ``datatype'' describes the type of data, which can take values like ``data table'', ``SEM image'' (from scanning electron microscope), ``cfu-plot'' (colony forming units, e.g. of bacteria), to give some examples. It corresponds to ``\href{http://purl.org/dc/terms/type}{dcterms.type}'';
	
	\item ``range'' is intended to detail a parameter range the resource is valid for. Examples might be a wavelength range (e.g. $400$ to $800$\,nm in case of emission spectra), or a magnification and an accelerating voltage in case of scanning electron microscope images;
	
	\item ``quality'' is considered to rank the level of scientific quality control. Allowed values are given by a controlled vocabulary consisting of ``verified'', ``published'', and ``reviewed''. Here, ``verified'' is the lowest quality level and means that the resource is checked for plausibility by the data creators and the data curators. ``published'' means that the data of the resource have already been published in a peer-reviewed paper. Finally, ``reviewed'' implies that the data resource has been peer-reviewed by an independent expert.
\end{enumerate}

Table~\ref{table:Plasma-MDS.resource} provides an overview of all qualifiers of the schema element ``plasma.resource''.
This schema element is mandatory because information on the available resources is crucial for finding and selecting relevant datasets. Note that this element must be provided for each resource if several digital objects are attached to the dataset.

\begin{table}[ht]
	\caption{Plasma-MDS metadata fields related to the resource.}
	\label{table:Plasma-MDS.resource}
	\small
	\centering   
	\begin{tabular}{ |l|P{2.5cm}|P{5cm}|P{1.5cm}|P{4cm}|}
		\hline
		\textit{\textbf{Id}} 
		& \textit{\textbf{Field}}         
		& \textit{\textbf{Definition}} 
		& \textit{\textbf{Format}}
		& \textit{\textbf{Example}} \\
		\hline
		5.1 & plasma.resource. filetype        
		& file type of the resource data
		& term list
		& csv, jpg, pdf\\
		\hline
		5.2 & plasma.resource. datatype        
		& kind of digital data which are saved with the resource
		& term list        
		& data table, SEM image, cfu-plot, high-speed video\\
		\hline        
		5.3 & plasma.resource. range        
		& range in which the resource is valid
		& free text
		& wavelength range: $400 \ldots 800$\,nm
		\\              
		\hline
		5.4 & plasma.resource. quality        
		& Data quality score
		& controlled list
		&  verified, published, reviewed\\
		\hline
	\end{tabular}
\end{table}

\subsection*{Examples}

To demonstrate how Plasma-MDS can be used for the annotation of research data, two examples from research in plasma technology are provided in the following. The first example comes from basic research studies of order phenomena in atmospheric pressure plasma jets and does not involve a plasma target.
The second example examines the origin of species in a liquid upon plasma interaction.
Free access to the Plasma-MDS metadata of both examples is provided by INPTDAT---a new  interdisciplinary data platform for plasma technology (\url{https://www.inptdat.de}, accessed: 2020-06-14). INPTDAT was build at INP with the aim to provide free and easy access to research data and information from all fields of plasma technology and plasma medicine. It aims to support the findability, accessibility, interoperability and reuse of data for the low-temperature plasma community.
Note that in the same way and in accordance with the approach shown in figure~\ref{fig:schema}, Plasma-MDS may be used for a unified extension of data catalogues in other plasma related domains.

%\subsubsection*{Correlation of helicality and rotation frequency of filaments in the ntAPPJ}

The first example demonstrates how Plasma-MDS was used for the annotation of digital data that has been used for analysis of the correlation of helicality and rotation frequency of filaments in a non-thermal atmospheric pressure plasma jet (ntAPPJ). 
Parts of the dataset have been pictured in figure 4 in Schäfer~\textit{et al.}~\cite{Schafer-2018-ID4871}. Tables~\ref{table:example1} and \ref{table:example1-resources} provide a preview of the Plasma-MDS metadata. Public access to the digital data and all metadata is provided by the dataset published with INPTDAT~\cite{example1}.

\begin{table}[ht]
	\caption{Preview of Plasma-MDS metadata for the dataset ``Correlation of helicality and rotation frequency of filaments in the ntAPPJ''~\cite{example1}. Full access to all metadata is provided by INPTDAT at \url{https://www.inptdat.de/node/43}.}
	\label{table:example1}
	\small
	\centering   
	\begin{tabular}{|l|P{4.5cm}|P{9.4cm}|}
		\hline
		\textit{\textbf{Id}} 
		& \textit{\textbf{Field}}         
		& \textit{\textbf{Value}}\\\hline
		\hline
		\multicolumn{3}{|l|}{\it Plasma source} \\
		\hline        
		1.1 & plasma.source.name
		& ntAPPJ, HF plasma jet
		\\\hline
		1.2 & plasma.source.application
		& PECVD
		\\\hline
		1.3 & plasma.source.specification
		& AC, high frequency, atmospheric pressure, non-thermal
		\\\hline   
		1.4 & plasma.source.properties
		& Non-thermal atmospheric pressure plasma jet (capacitively coupled) operated in a self-organized regime (locked mode). Power: 7 to 9 W; Frequency: 27.12 MHz; Flow rate: 400 to 800 sccm argon; Electrodes: ring configuration, distance 5 mm, width 5 mm; Capillary: inner diameter 4 mm, outer diameter 6 mm
		\\\hline                 
		1.5 & plasma.source.procedure        
		& The measurements occur 30 minutes after temperature conditioning of the plasma source for each parameter setting.
		\\\hline
		\hline
		\multicolumn{3}{|l|}{\it Plasma medium} \\
		\hline
		2.1 & plasma.medium.name
		& Ar
		\\\hline   
		2.2 & plasma.medium.properties
		& Flowrate: 0.4 to 0.8 slm; Pressure: 1 bar; Temperature: 300 to 1000 K; Purity: argon 6.0
		\\\hline                 
		2.3 & plasma.medium.procedure        
		& Standard conditions of the argon gas are assured.
		\\\hline
		\hline
		\multicolumn{3}{|l|}{\it Plasma target} \\
		\hline        
		3.1 & plasma.target.name
		& --
		\\\hline   
		3.2 & plasma.target.properties
		& --
		\\\hline                 
		3.3 & plasma.target.procedure        
		& --
		\\\hline
		\hline
		\multicolumn{3}{|l|}{\it Diagnostics} \\
		\hline            
		4.1 & plasma.diagnostics.name
		& laser schlieren deflectometry, fast imaging
		\\\hline   
		4.2 & plasma.diagnostics.properties
		& The filament behaviour has been visualized optically by means of imaging. The LSD set-up consists of a He-Ne laser (Linos) and a high-speed CMOS camera (Photon Focus). \newline
		$\dots$
		%The image sensor has 1312$\times$1082 pixels, pixel size 8\,$\upmu$m. The laser beam is classified as 3 A, 632.82 nm, cw. The output power of the laser is 0.6 W. The laser beam is specified by the spot diameter of 490\,$\upmu$m and beam divergence of 1.7 mrad. The parameters of the beam enable to characterize filament structures with diameters down to approx. 100\,$\upmu$m.
		\\\hline  
		4.3 & plasma.diagnostics.procedure
		& The displacement of the laser spot on the image sensor of the camera is monitored directly. The laser beam is led perpendicular through the effluent region, close to the nozzle of the capillary (z\,=\,0\,mm). The output power of the laser is \newline
		$\dots$
		%sufficiently low to avert any influence of the beam on the properties of the plasma jet. Deflection of the laser beam due to refraction in the gradient field of the refractive index is recorded at a distance of 80.6 cm from the plasma jet at 10 kHz sample frequency.
		\\\hline   		           
	\end{tabular}
\end{table}

\begin{table}[ht]
	\caption{Preview of Plasma-MDS resource metadata for the dataset ``Correlation of helicality and rotation frequency of filaments in the ntAPPJ''~\cite{example1}. Full access to all resource metadata is provided by INPTDAT at \url{https://www.inptdat.de/node/84}, \url{https://www.inptdat.de/node/85}, \url{https://www.inptdat.de/node/86}, \url{https://www.inptdat.de/node/87}.}
	\label{table:example1-resources}
	\small
	\centering   
	\begin{tabular}{|l|P{4cm}|P{2.1cm}|P{2.1cm}|P{2.1cm}|P{2.1cm}|}
		\hline
		\textit{\textbf{Id}} 
		& \textit{\textbf{Field}}         
		& \multicolumn{4}{|c|}{\textit{\textbf{Value}}}
		\\\hline\hline 
		\multicolumn{2}{|c|}{} & \it Resource 1 & \it Resource 2 & \it Resource 3 & \it Resource 4\\\hline
		5.1 & plasma.resource.filetype
		& png & csv & csv & csv
		\\\hline   
		5.2 & plasma.resource.datatype
		& high-speed image & data table & data table & data table
		\\\hline
		5.3 & plasma.resource.range
		& Power: 7 to 9\,W; Gas flow rate: 0.4 to 0.7\,slm
		&  Power: 7\,W; Rotation frequency range: 0 to 90\,Hz
		& Power: 8\,W; Rotation frequency range: 0 to 90\,Hz
		& Power: 9\,W; Rotation frequency range: 0 to 90\,Hz
		\\\hline   
		5.4 & plasma.resource.quality
		& verified & published & published & published
		\\\hline              
	\end{tabular}
\end{table}       

%\subsubsection*{Non-thermal plasma in contact with water: the origin of species}
In the second example, Plasma-MDS was used for the annotation of a dataset published in the York Research Database (\url{https://pure.york.ac.uk/portal/}, accessed: 2020-06-14). This shows how Plasma-MDS and the research data platform INPTDAT can be used to improve the findability and reusability of digital research data published elsewhere. The dataset published with INPTDAT at \url{https://www.inptdat.de/node/98} includes all Plasma-MDS metadata and refers to the original dataset published in the York Research Database~\cite{example2}. In this case, detailed information on plasma source, plasma medium, plasma target, applied diagnostics, and published resources were extracted from the journal article and its supporting information~\cite{Gorbanev-2016} to which the digital data belong. Tables~\ref{table:example2} and \ref{table:example2-resources} provide a preview of the Plasma-MDS metadata.
\begin{table}[ht]
	\caption{Preview of Plasma-MDS metadata for the dataset ``Non-thermal plasma in contact with water: the origin of species''~\cite{example2}. Full access to all metadata is provided by INPTDAT at \url{https://www.inptdat.de/node/98}.}
	\label{table:example2}
	\small
	\centering   
	\begin{tabular}{|l|P{4.5cm}|P{9.4cm}|}
		\hline
		\textit{\textbf{Id}} 
		& \textit{\textbf{Field}}         
		& \textit{\textbf{Value}}\\\hline
		\hline
		\multicolumn{3}{|l|}{\it Plasma source} \\
		\hline        
		1.1 & plasma.source.name
		& kHz plasma jet
		\\\hline
		1.2 & plasma.source.application
		& reactive species generation
		\\\hline
		1.3 & plasma.source.specification
		& AC, low frequency, atmospheric pressure, non-thermal
		\\\hline   
		1.4 & plasma.source.properties
		& The plasma was ignited in a quartz tube (4 mm ID and 6\,mm OD, 100\,mm length) surrounded by copper electrodes (10\,mm width) separated by 20\,mm. A PVM500 Plasma \newline
		$\dots$
		%Resonant and Dielectric Barrier Corona Driver power supply (Information Unlimited) was used to sustain the plasma. The distance between the electrodes was 20\,mm in all experiments. Voltage and frequency were kept constant throughout all experiments at 18.3 $\pm$ 0.2\,kV (peak-to-peak) and 24.9\,kHz, respectively. The return current values were between ca. 4 and 7\,mA. The experimental setup was positioned inside a large Faraday cage with the mesh size of 22\,mm.
		\\\hline                 
		1.5 & plasma.source.procedure        
		& In a typical experiment, 100\,$\upmu$L of liquid sample was placed in a well on top of a glass stand inside the reactor. The distance from the nozzle to the sample was 10\,mm unless \newline
		$\dots$
		%stated otherwise. In experiments when the samples were at the 4\,mm distance from the sample to the nozzle, the distance between the live electrode and the sample was maintained at 20\,mm. Thus, the plasma length from the core plasma remained the same throughout all experiments, and the ratio of its quartz surroundings changed. The reactor was flushed with the feed gas for 20\,s and then exposed to plasma for 60\,s.
		\\\hline
		\hline
		\multicolumn{3}{|l|}{\it Plasma medium} \\
		\hline
		2.1 & plasma.medium.name
		& He, H$_2$O
		\\\hline   
		2.2 & plasma.medium.properties
		& The plasma was operated with a feed gas of helium with oxygen and water admixtures controlled by mass flow controllers (MFCs) (Brooks Instruments and Brooks\newline
		$\dots$         
		% Instruments 0254 microcomputer controller). All experiments were carried out with a total flow of feed gas of 2 L/min. Helium He (A Grade, 99.996\,\%) and oxygen O$_2$ (Zero Grade, 99.6\,\%) were supplied by BOC UK. All chemicals were used as received.
		\\\hline                 
		2.3 & plasma.medium.procedure        
		& The experiments involving different feed gas humidity were performed by using split helium flow (i.e., by mixing dry helium with water-saturated helium in desired proportions). \newline
		$\dots$
		% Water-saturated helium was made by bubbling dry helium through a water-filled Drechsel flask at 20\,$^{\circ}$C. The relative humidity was determined by weighing the flask before and after the experiment and comparing the data with the available literature values.
		\\\hline
		\hline
		\multicolumn{3}{|l|}{\it Plasma target} \\
		\hline        
		3.1 & plasma.target.name
		& H$_2$O$_2$, H$_2$SO$_4$, NaN$_3$, D$_2$O, PBN, TEMP, TEMPO, sodium tosylate, H$_2$O, DMPO, DEPMPO, potassium bis(oxalato)oxotitanate(IV) dihydrate
		\\\hline   
		3.2 & plasma.target.properties
		& Hydrogen peroxide H$_2$O$_2$ (30\,\%), sulphuric acid H$_2$SO$_4$ ($>$95\,\%) and sodium azide NaN$_3$ ($\geq$99.5\,\%) were purchased from Fluka. Deuterium oxide D$_2$O (99.9 atom\,\% D), \newline
		$\dots$ 
		%N-tert-butyl-$\alpha$-phenylnitrone (PBN) (98\,\%), 2,2,6,6-tetramethylpiperidine (TEMP) ($\geq$99\,\%), 2,2,6,6-tetramethylpiperidine 1-oxyl (TEMPO) (98\,\%), sodium p-toluenesulfonate (sodium tosylate) (95\,\%), cinnamoyl chloride (98\,\%) and H$_2^{18}$O (97\,\%) were obtained from Sigma-Aldrich. 5,5-Dimethyl-1-pyrroline N-oxide (DMPO) ($\geq$99\,\%) and 5-diethoxyphosphoryl-5-methyl-1-pyrroline N-oxide (DEPMPO) ($\geq$99\,\%) were purchased from Dojindo Molecular Technologies, Inc. and Enzo Life Sciences, respectively. Potassium bis(oxalato)oxotitanate(IV) dihydrate was obtained from Alfa Aesar. H$_2^{17}$O was purchased from Icon Isotopes. De-ionised water was used for the preparation of the solutions. All chemicals were used as received.
		\\\hline                 
		3.3 & plasma.target.procedure        
		& In spin trapping experiments, a 100\,mM solution of a spin trap (PBN, DMPO or DEPMPO) was prepared in H$_2$O, H$_2^{17}$O or D$_2$O. Ozone was measured in 60\,mM aqueous \newline
		$\dots$
		%solutions of TEMP (sodium azide was added in concentrations of 100\,mM where stated). In control experiments, solutions of each spin trap were treated with the plasma for the periods of 15, 30, 45 and 60\,s.
		\\\hline
		\hline
		\multicolumn{3}{|l|}{\it Diagnostics} \\
		\hline            
		4.1 & plasma.diagnostics.name
		& spin-trapping, isotopic labelling, EPR spectroscopy, OES
		\\\hline   
		4.2 & plasma.diagnostics.properties
		& A high voltage probe (Tektronix P6015A) and current probe (Ion Physics Corporation CM-100-L) were used with a Teledyne LeCroy WaveJet 354A oscilloscope to measure time  \newline
		$\dots$ 
		%resolved current and voltage. OES measurements of the plasma between the electrodes were performed with Ocean Optics HR-4000CG-UV-NIR spectrophotometer. Electron paramagnetic resonance (EPR) measurements were carried out on a Bruker EMX Micro EPR spectrometer. The EPR analysis parameters were as follows: frequency 9.83\,GHz, power 3.17\,mW, modulation frequency 100\,kHz, modulation amplitude 1\,G, time constant 40.96\,msec, number of scans 5, sweep width 100\,G (DMPO and PBN adducts, TEMPO) or 170\,G (DEPMPO addcuts). EPR spectra simulations were performed on NIH P.E.S.T. WinSIM software ver. 0.96. Concentration of H₂O₂ in the samples was determined by UV-Vis measurements performed on a UV-1800 Shimadzu UV-Vis Spectrophotometer with Optical Glass High Precision Cells (10 mm light path) provided by Hellma Analytics.
		\\\hline   
		4.3 & plasma.diagnostics.procedure
		& UV-Vis calibration was done using 500\,$\upmu$L titanium(IV) reagent with added 300\,$\upmu$L aqueous hydrogen peroxide solutions in a range of concentrations 0.0979--4.895 mM. \newline
		$\dots$
		%UV-Vis spectra of samples were recorded by adding a mixture of 65 µL of plasma-exposed sample (taken immediately after plasma exposure) with 235 µL of H2O to 500 µL of titanium(IV) reagent. The resulting solutions were incubated for 1 min before analysis. The H₂O₂ concentration was determined from the UV-Vis intensity of the peak at 400 nm.
		\\\hline  		          
	\end{tabular}
\end{table}
\begin{table}[ht]
	\caption{Preview of Plasma-MDS resource metadata for the dataset ``Non-thermal plasma in contact with water: the origin of species''~\cite{example2}. Full access to all resource metadata is provided by INPTDAT at \url{https://www.inptdat.de/node/99}.}
	\label{table:example2-resources}
	\small
	\centering   
	\begin{tabular}{|l|P{4cm}|P{8.4cm}|}
		\hline
		\textit{\textbf{Id}} 
		& \textit{\textbf{Field}}         
		& \textit{\textbf{Value}}
		\\\hline
		5.1 & plasma.resource.filetype
		&  html
		\\\hline   
		5.2 & plasma.resource.datatype
		&  external resource
		\\\hline
		5.3 & plasma.resource.range
		& The results of the plasma exposure of the samples (e.g., the absolute values of concentration of DMPO-OH) were largely affected by small changes in the configuration of the jet, such as the electrodes contact with the quartz tube, the depth of the tube protrusion inside the reactor, and the vertical alignment of the tube. However, while the numerical values changed, the observed trends remained persistent. For example, the concentration of DMPO-OH increased with the initial introduction of H$_2$O to He feed gas and decreased with higher H$_2$O content, the concentration of DMPO-OH was lower at 4\,mm distance than 10\,mm, etc. Thus, the error assessment was performed within a set configuration of the jet for several conditions. Conditions of less uniform plasma nature (i.e., in the presence of large amounts of admixtures in the feed gas) generally lead to an increase in standard deviation of the concentration values. The maximum deviation from the mean was found to be ca. 12\,\%.
		\\\hline   
		5.4 & plasma.resource.quality
		&  published
		\\\hline              
	\end{tabular}
\end{table}       
Here, INPTDAT does not provide direct access to the digital object, i.e. the research data but strongly enhances the findability and reusability of the original data by means of Plasma-MDS.
\clearpage

\section*{Discussion}

The plasma metadata schema Plasma-MDS was developed to complement basic metadata schemata with metadata fields for the collection of domain-specific information. This was demonstrated by means of two examples (Tables~\ref{table:example1}--\ref{table:example2-resources}). This section discusses how Plasma-MDS supports the transfer of FAIR data principles (Table \ref{table:FAIR}) into practice to enable data driven plasma science.

\paragraph*{Findability:} 

To be findable, machine readable metadata should allow the discovery of relevant datasets by humans and computer systems~\cite{snf}.
According to the FAIR data principles (Table~\ref{table:FAIR}), a globally unique and persistent identifier should be assigned to each dataset. This identifier allows to find, track and cite data and their metadata.
Plasma-MDS makes use of metadata fields for the unique identifier as part of basic metadata schemata (e.g. Dublin Core metadata field ``\href{http://purl.org/dc/terms/identifier}{dcterms.identifier}''). 
Furthermore, Plasma-MDS strongly supports findability by providing specific fields to collect rich domain-specific metadata. The metadata collected by Plasma-MDS aims to allow researchers to properly understand the nature of the dataset by including descriptive information about the context, conditions, and quality of the data as demonstrated by the two examples in Tables~\ref{table:example1}--\ref{table:example2-resources}. Particularly, the metadata fields using controlled lists (plasma.source.specification and plasma.resource.quality) and term lists (e.g. qualifier ``name'' for all schema elements) also support the automated processing of metadata by computer systems. The collection of general information (e.g. creator, date, and license) is already supported by basic metadata schemata (Dublin Core, DataCite Metadata Schema, and others). To take full advantage of the benefits of Plasma-MDS, the implementation of Plasma-MDS in a (meta)data repository is needed. However, generic or institutional data repositories usually do not provide the possibility to collect domain-specific information on deposited datasets. INPTDAT is the first data platform that implements Plasma-MDS and indexes the domain-specific metadata to provide elaborate search features for interdisciplinary datasets in the field of plasma technology. The Schema.org representation of the DataCite Metadata Schema~\cite{Fenner2017} is used by INPTDAT for registration of digital object identifiers (DOI) and as an important means for increasing the findability of datasets by major search engines like Google, Microsoft, Yahoo and Yandex~\cite{Sansone2017}.

\paragraph*{Accessibility:} 

To be accessible, data and metadata should be stored for the long term such that they can be easily accessed by humans and computer systems using standard communication protocols~\cite{snf}. 
This requirement cannot be met by the metadata schema itself, but by the repository in which the (meta)data is stored. Any (meta)data repository providing Plasma-MDS as metadata schema should meet the requirements for accessibility according to the FAIR data principles (Table~\ref{table:FAIR}). Therefore, the data platform INPTDAT uses public APIs (application programming interfaces) to provide open access to general as well as domain-specific metadata in different formats. No authentication and authorization of users is required to access metadata. Furthermore, INPTDAT maintains all metadata physically separated from data files and provides the possibility to easily extract and move metadata to other repositories by public APIs.

\paragraph*{Interoperability:}

To be interoperable, data should be ready to be exchanged, interpreted, and combined in a (semi)automated way with other datasets by humans and computer systems~\cite{snf}. Therefore, community standards for data management and, in particular, established vocabularies/ontologies/thesauri are required. In this respect, Plasma-MDS can be seen as a first step towards increasing awareness and the development of a common standard. Appropriate collaborative structures are to be set up that allow to develop, maintain, and document controlled vocabularies that themselves again fulfil the FAIR data principles (Table~\ref{table:FAIR}). Where appropriate, this should build on established ontological resources in related areas. 
Finally, the possibility to include qualified references to other (meta)data is provided by basic metadata schemata (e.g. Dublin Core metadata field ``\href{http://purl.org/dc/terms/relation}{dcterms.relation}'').

\paragraph*{Reusability:}

To be reusable,  the provided metadata must ensure that the dataset can be used in future research and that it can be integrated with other compatible data sources. The conditions under which the data can be reused should be clear to humans as well as computer systems~\cite{snf}.
Therefore, the FAIR data principles (Table~\ref{table:FAIR}) demand a detailed description of the dataset including information on what the dataset contains, how it was generated and processed, and the conditions under which the data can be reused.
Future use of Plasma-MDS will turn out whether the implemented qualifiers for the schema elements ``plasma.source'', ``plasma.medium'', ``plasma.target'', ``plasma.diagnostics'', and ``plasma.resource'' suffice to  achieve this requirement and which adjustments might be necessary in revised versions of Plasma-MDS. It is worth mentioning that each metadata schema is not fixed but subject to regular updates.
Furthermore, it is important to note that metadata schemata are only as good as they are being used. It is the responsibility of users to provide the required information with their data. 
Basic metadata schemata already provide fields to include information on the data usage license (e.g. Dublin Core metadata field ``\href{http://purl.org/dc/terms/rights}{dcterms.rights}''). A discussion of appropriate licenses for data publications is beyond the scope of this paper. In general, the Creative Commons (CC) license CC~BY~4.0 (\url{https://creativecommons.org/licenses/by/4.0/}, accessed: 2020-06-14) is recommendable for research data publications. More information and recommendations on how to license research data are provided, e.g. by the Digital Curation Centre~\cite{Ball-2014}.
The association of data with detailed provenance, appears to be the most challenging FAIR data principle (Table~\ref{table:FAIR}). At the same time, it is the most important precondition for reliable reuse of research data. Detailed information about the provenance of data allows researchers to understand how the data were generated, in which context it can be reused, and how reliable it is. 
With the qualifiers ``properties'' and ``procedure'' for the different schema elements, Plasma-MDS is prepared to collect the relevant metadata. However, it is difficult to ensure that third parties will be able to fully understand and reproduce the workflow of data creation, especially for the large number of experimental methods in the field of low-temperature plasmas for which no standard operation procedures (SOPs) are available yet. In this respect, Plasma-MDS may give the impetus to agree on certain SOPs and data annotation standards for widely used experimental methods in the low-temperature plasma community. Finally, data and metadata should meet domain-relevant community standards. Obviously, this requirement is only applicable if community standards or best practices for data archiving and sharing exist. Plasma-MDS can possibly contribute to the establishment of such standards.

\paragraph*{Outlook:}

In conclusion, there is still a need for action, particularly in the introduction of community standards with respect to controlled terminologies. 
Few sub-areas of plasma science and technology could benefit from already established vocabularies and ontologies, e.g. in biomedical science (e.g. OBI, see \url{http://obi-ontology.org}, accessed: 2020-06-14) (relevant for plasma medicine) and material science (e.g. EMMO, see \url{https://emmc.info/emmo-info/}, accessed: 2020-06-14) (relevant for plasma surface technology).
In this regard there is the desire that Plasma-MDS will be integrated into other data repositories and further developed by the plasma community, particularly with respect to controlled terminologies.
The establishment of research data management standards in widespread plasma research areas is seen as a basic prerequisite for extensive data-driven research and will be followed up, e.g. within the framework of the initiative on data driven plasma science.
The long-term goal is to establish Plasma-MDS as a widespread community standard, which supports the reuse of research data and promotes data driven plasma science up to the point where research data management becomes everyday practice in the plasma community.

\section*{Methods}

Plasma-MDS has been developed given practical requirements of research in plasma technology.
In accordance with typical plasma processes and applications, the schema contains metadata fields to collect annotations about the plasma source, plasma medium, and plasma target involved in the study from which research data were obtained. Furthermore, metadata for the respectively applied diagnostics and modelling/simulation methods are collected. Finally, resource metadata fields are included to describe the individual digital objects belonging to the dataset. The following approaches have been used to compile the five schema elements of Plasma-MDS:

\begin{enumerate}
	\item Scientist interviews and use cases. A number of interviews has been performed with scientists, which are active in different fields of applied plasma physics and plasma medicine. This was important to broaden the expert knowledge of the authors and to identify use cases which gave impact on required information that is needed to ensure optimal findability and reuse of heterogeneous data in the field of research.
	
	\item Exploration of existing metadata schemata. Different metadata schemata have been investigated to include existing standards like Dublin Core and DataCite, and to match to metadata fields of existing software systems like DSpace, CKAN and DKAN. Domain specific metadata schemata, like HEPdata~\cite{Maguire_2017} and the ETHZ metadata schema (\url{https://documentation.library.ethz.ch/display/RC/Metadata+Schema}, accessed: 2020-06-14) gave rise to structure Plasma-MDS according to the requirements of the addressed scientific community.
	
	\item Expert workshops. Several workshops have been performed to present and discuss the state of development of Plasma-MDS metadata schema. Workshops acted experts audits in the domain of applied plasma physics and plasma medicine as well as in the field of research data management. The workshops identified missing ``metadata'' and refined controlled lists.
	
	\item Keywording of publications. Several publications in the field of applied plasma physics and plasma medicine have been enriched with metadata by test to ensure the proper choice of metadata schema element and qualifier. By this means Plasma-MDS has been checked for consistency and completeness.
\end{enumerate}

Having the basic schema elements and their attributes fixed, JSON Schema (\url{https://json-schema.org}, accessed: 2020-06-14) was used for formal representation of Plasma-MDS. The JSON representation of Plasma-MDS is retrievable at \url{https://www.plasma-mds.org/schema/plasma-mds.json}.
Note that the JSON schema file may also serve for implementation of Plasma-MDS in third-party systems and validation of instance files against the schema using existing JSON validation tools, e.g. Ajv: Another JSON Schema Validator (\url{https://ajv.js.org}, accessed: 2020-06-14) or JSON Schema Validator (\url{https://www.jsonschemavalidator.net}, accessed: 2020-06-14), respectively.

\section*{Data availability}
Archival copies of the metadata records for the presented examples as well as the JSON representation of Plasma-MDS are available in the ``Zenodo'' repository~\cite{zenodo-archive}.  

\section*{Code availability}
DKAN Open Data Platform (\url{https://getdkan.org}) is used as a basis the for implementation of the INPTDAT platform. The INPTDAT code base particularly includes the implementation of Plasma-MDS and is publicly available at GitHub (\url{https://github.com/markus-m-becker/INPTDAT}). 
INPTDAT is freely-available under the "GNU General Public License, version 2 or any later version" license in agreement with the DKAN licensing.

%DKAN is licensed under GPLv2 or later. The DKAN license also covers the used DKAN modules such as Open Data Schema Map and the taxonomy features, see \url{https://docs.getdkan.com/en/latest/license.html}. The source code is publicly available at GitHub (\url{https://github.com/GetDKAN/dkan}).

%\bibliography{references}


\section*{Acknowledgements}

M.M.B thanks D. Loffhagen and D. Uhrlandt for supportive discussions and the scientists involved at INP for helpful suggestions for the design of the metadata schema. 
The work was funded by the Federal Ministry of Education and Research (BMBF) under the grant marks 16FDM005 and 16QK03A. 
The responsibility for the content of this publication lies with the authors.

\section*{Author contributions}

St.F., L.P. and M.M.B. designed the research and discussed with all other co-authors. J.S. and D.O'C. contributed the example applications of Plasma-MDS. All authors gave iterative feedback on the metadata schema as well as the manuscript. St.F. and M.M.B. wrote the manuscript, with contributions from the other co-authors. All co-authors have contributed to its final version.

\section*{Competing interests}

The authors declare no competing interests.

\end{document}